# Generation of Entangled Ancilla States for use in
# Linear Optics Quantum Computing

J.D. Franson, M. M. Donegan, and B.C. Jacobs
Johns Hopkins University
Applied Physics Laboratory
Laurel, MD 20723

*Abstract:*

Quantum logic operations can be performed using linear optical elements, additional photons (ancilla), and post-selection based on measurements made on the ancilla. Here we describe a method for generating the required entangled state of $n$ ancilla photons using elementary logic gates and post-selection. This approach is capable of generating the ancilla states required for either the original proposal by Knill, Laflamme, and Milburn [Nature **409**, 46 (2001)] or those required for the more general high-fidelity approach [J. D. Franson, M. M. Donegan, M. J. Fitch, B. C. Jacobs, and T. B. Pittman, Phys. Rev. Lett. **89**, 137901 (2002)]. We also show that the entangled ancilla photons could be generated using a series of quantum wells coupled with tunnel junctions.

## I. INTRODUCTION

Quantum logic operations can be performed using linear optical elements, such as beam splitters and phase shifters, combined with measurements made on a set of $n$ entangled ancilla photons. In the original approach suggested by Knill, Laflamme, and Milburn (KLM) [1], post-selection can be used to obtain the correct logical output with a failure rate that scales as $2/(n+1)$. We have subsequently shown [2] that the entangled ancilla state can be chosen in such a way as to maximize the fidelity of the output with an error rate that scales as $4/(n+1)^2$. In either case, a practical implementation of linear optics quantum computing will require a method for generating the appropriate entangled ancilla states with a sufficiently large value of $n$.

Here we describe two different methods for generating ancilla photons in the required entangled states. The first method makes use of elementary logic gates based on post-selection as suggested by KLM. We describe a more general algorithm that can generate the entangled ancilla states required by either the original KLM approach [1] or the high-fidelity approach [2]. Although the post-selection method is relatively straightforward, its efficiency decreases exponentially with $n$. We show that the same resources are required to generate the ancilla for the original KLM approach or for the high-fidelity approach.

As a potential solution to the problem of exponentially-increasing resources, we also describe a method for generating the required entangled states using a series of quantum wells and the Coulomb blockade effect [3, 4]. By applying the proper voltages to a set of tunnel barriers connecting the quantum wells, it is possible to generate the corresponding entangled state of $n$ excited electrons, which can then emit the ancilla photons into a set of single-mode optical fibers [5]. The potential advantage of this approach is that the required number of operations scales as $n^2$ rather than exponentially.

The feasibility of these approaches will depend on the availability of high-efficiency single-photon sources and detectors, as well as efficient coupling between optical fibers and bulk components. As a result, it may be desirable to consider the possibility of alternative approaches that minimize or eliminate the need for large numbers of ancilla photons and high-efficiency detectors [6].

We begin in Section II by reviewing the entangled ancilla states required for the KLM approach as well as the high-fidelity approach. The generation of the ancilla photons using elementary logic gates and post-selection is discussed in Section III. The use of quantum wells to generate the entangled ancilla states is then described in Section IV. A summary and conclusions are presented in Section V.

## II. REQUIRED ANCILLA STATES

The KLM and high-fidelity approaches are both based on an efficient method of quantum teleportation [7, 8]. It is well known that quantum teleportation requires a Bell-state measurement, which cannot be done with certainty using linear optical devices [9, 10]. KLM showed [1] that quantum teleportation could, however, be performed probabilistically using an entangled state of $n$ photons that are distributed between two sets of $n$ optical modes, which we refer to as registers x and y as illustrated in the top half of Figure 1. Each register could consist of $n$ single-mode optical fibers, for example, with one set leading to the left and the other to the right. This set of $n$ entangled photons replaces the single pair of entangled photons that is conventionally used for teleportation.

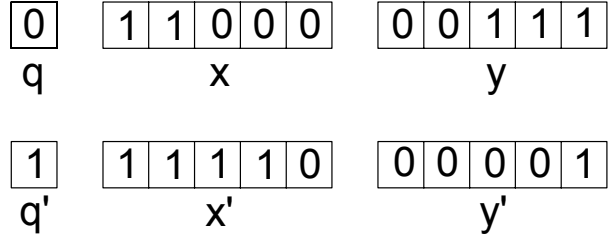

Figure 1. Ancilla registers x and y used in the quantum teleportation of qubit q. Each mode initially contains 0 or 1 photons, while the total number of photons in x and y has a fixed value $n$ ($n = 5$ for the example shown here). The index j in the entangled state of Eq. (1) corresponds to the number of photons in x. Similar ancilla registers x' and y' are used to teleport a second qubit q' during logic operations.

Aside from a normalization constant, the entangled state of the ancilla in registers x and y required for the teleportation of a single qubit is given by

$$|\psi_A\rangle = \sum_{j=0}^{n} f(j) \, |1\rangle^j \, |0\rangle^{n-j} \, |0\rangle^j \, |1\rangle^{n-j} \tag{1}$$

Here $n$ is an arbitrary integer that corresponds to the total number of ancilla photons while j is the number of those photons in register x. The first factor of $|1\rangle^j$ denotes one photon in each of the first j modes of register x, the factor of $|0\rangle^{n-j}$ represents no photons in the next $n-j$ modes of x, the second factor of $|0\rangle^j$ represents no photons in the first j modes of register y, and the final factor of $|1\rangle^{n-j}$ denotes a single photon in each of the remaining modes of register y. The value of the function $f(j)$ for the high-fidelity approach is given in Ref. [2], while it has a constant value of 1 in the original KLM approach [1].

The teleportation of a single qubit involves a quantum Fourier transform [11] applied to registers q and x, followed by a measurement of the number of photons in each of those registers [1, 2]. The output of the teleportation process is then selected from one of the $y$ registers, based on the results of the measurements. Feed-forward control [12] in the form of a phase shift determined by the results of the measurements may also be required.

A controlled sign gate (a controlled Pauli operator $\hat{Z}$) can be implemented by teleporting two qubits q and q' using two sets of entangled ancilla registers, x, y, x', and y', as illustrated in Figure 1. The two qubits are teleported independently, but the two sets of ancilla are now assumed to be in an entangled state of the form

$$|\psi_{AA'}\rangle = \sum_{j=0}^{n} f(j)\,|1\rangle^{j}\,|0\rangle^{n-j}\,|0\rangle^{j}\,|1\rangle^{n-j} \sum_{j'=0}^{n} (-1)^{jj'} f(j')\,|1\rangle^{j'}\,|0\rangle^{n-j'}\,|0\rangle^{j'}\,|1\rangle^{n-j'} \qquad (2)$$

Here $n$ is an arbitrary integer as in Eq. (1) while $j$ and $j'$ correspond to the number of photons in registers x and x', respectively. The notation is the same as in Eq. (1), except that the unprimed modes refer to registers x and y, while the primed modes refer to registers x' and y'. A controlled-NOT gate can be constructed using such a controlled sign gate and single-qubit operations.

It can be seen that the ancilla state of Eq. (2) is the tensor product of two independent states of the form given in Eq. (1), aside from the factor of $(-1)^{jj'}$ which entangles the two. Implementing this phase factor complicates the generation of the required ancilla state, as is described in the next two sections.

## III.    GENERATION OF ANCILLA USING POST-SELECTION

A variety of elementary quantum logic gates [1, 13-19], including a controlled-NOT, can be implemented using at most two ancilla photons. These devices succeed with probabilities ranging from 1/16 to ¼ using one ancilla that consist of either single photons or the two-photon Bell-state $|\Phi^{+}\rangle$ produced by parametric down-conversion [20-22]. Since a controlled-NOT gate and single-qubit operations are universal, it follows that these elementary logic gates can be used to generate any desired state of $n$ ancilla photons starting from $n$ single photons. This is equivalent to a post-selection process in which a large number of single photons or $|\Phi^{+}\rangle$ states are used as a resource to generate the desired ancilla state of $n$ photons.

Here we describe a post-selection process of this kind that can generate the entangled ancilla state required for either the original KLM or the high-fidelity approach. The generation of the entangled ancilla state can be performed in three steps:   (a) Generation of a single photon in each mode of registers y and y'. (b) Conditional transfer of single photons from register y to x and from y' to x'. (c) Entanglement of the two sets of registers by applying a phase shift of $(-1)^{jj'}$. Each of these steps will now be described in more detail.

### A. Generation of single photons in y and y'

The first step in the generation of the required ancilla state is to transfer a single photon into each mode of the y and y' registers, as illustrated in Fig. 2. (It is assumed that all of the registers were in the vacuum state initially.) A number of single-photon sources have been experimentally demonstrated [23-26]. If each mode of the two registers consists of an optical fiber, then the photons can simply be routed from the single-photon sources into the appropriate fibers.

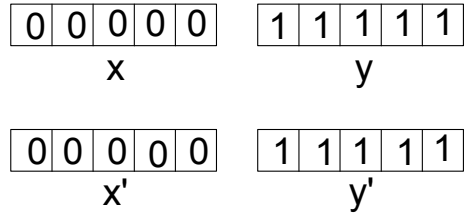

Figure 2. The first step in the generation of the entangled ancilla state is the transfer of a single photon into each mode of registers y and y'.

An error in the subsequent logic operations will occur if the single-photon source fails to generate a photon as expected. As a result, the efficiency of the single-photon source is an essential requirement for fault-tolerant quantum computing. An ideal way to achieve high efficiencies would be to generate a pair of photons directly in an optical fiber using down-conversion [27], which would avoid any coupling losses. The detection of one member of the pair would then ensure that the other photon is present in the fiber with high probability [26].

### B. Conditional transfer of photons

The next step in the generation of the entangled ancilla consists of conditional transfers of photons from register y to x and from register y' to x' in order to create two independent ancilla of the form shown in Eq. (1). For example, we first need to transfer the photon initially in the left-most mode of y into the left-most mode of x with the appropriate probability amplitude, as illustrated in Fig. 3. This process must create a coherent superposition of the two states, but it will be convenient to describe the process in terms of the probability $P_1$ that the photon would be found in x after the process has been completed. The required value of $P_1$ will be specified shortly.

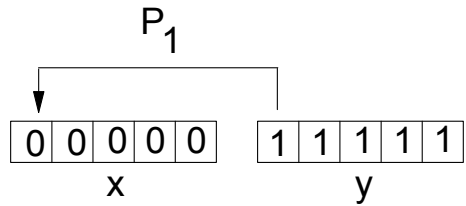

Figure 3. Coherent transfer of a single photon from register y to the appropriate mode of register x with probability $P_1$.

Once the first photon has been transferred with probability $P_1$, a second photon must be transferred with probability $P_2$ from the second mode from the left-hand-side of register y to the corresponding mode in register x, as illustrated in Figure 4. This process is conditional, however, in the sense that it is to be performed only if the first transfer was successful, i.e., only if the first mode of register x actually contains a photon. Once again, this process, if applied, produces a coherent superposition state, although it is convenient to describe it in terms of the corresponding probability $P_2$ that the second photon would be found in x.

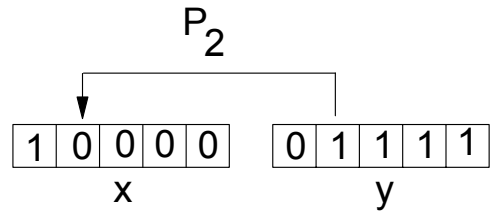

Figure 4. Conditional transfer of a photon from one mode of register y to the corresponding mode of register x. This process only takes place if the first mode in register x contains a photon, in which case it creates a superposition of states in which the second photon would be found in register x with probability $P_2$.

Conditional photon transfers of this kind can be implemented using the interferometer shown in Figure 5. Here the solid lines represent mirrors, the dashed lines represent 50/50 beam splitters, the modes have been labeled a through f, and $\phi$ is a phase shift of either $0^0$ or $180^0$. If a photon is initially present in path a, the initial state of the system is

$$|\psi\rangle = \hat{a}^\dagger |0\rangle \qquad (3)$$

Here the operators $\hat{a}^\dagger$, $\hat{b}^\dagger$, etc., create a single photon in the corresponding mode, as usual. The first beam splitter applies the operator transformation

$$\hat{a}^\dagger \rightarrow T\hat{d}^\dagger + iR\hat{c}^\dagger$$
$$\hat{b}^\dagger \rightarrow T\hat{c}^\dagger + iR\hat{d}^\dagger \qquad (4)$$

where R and T are the reflection and transmission coefficients of the beam splitter (both real numbers). The phase shifter applies the transformation

$$\hat{d}^\dagger \rightarrow \exp[i\phi]\hat{d}^\dagger \qquad (5)$$

after which the second beam splitter applies the transformation

$$c^\dagger \rightarrow T\hat{f}^\dagger + iR\hat{e}^\dagger$$
$$\hat{a}^\dagger \rightarrow T\hat{e}^\dagger + iR\hat{f}^\dagger \tag{6}$$

After these transformations the output state reduces to

$$|\psi_{out}\rangle = -\hat{e}^\dagger|0\rangle \quad for \quad \phi = 180^0$$
$$|\psi_{out}\rangle = [-(1-2T^2)\hat{e}^\dagger + 2iRT\hat{f}^\dagger]|0\rangle \quad for \quad \phi = 0^0 \tag{7}$$

It can be seen that the incident photon always appears in the output mode e if $\phi = 180^0$, while it is transferred to mode f with probability $4R^2T^2 = 4(1-T^2)T^2$ if $\phi = 0^0$. By choosing the proper value of T, it is thus possible to transfer the photon with any desired probability P provided that $\phi = 0^0$. An appropriate phase shift must also be applied.

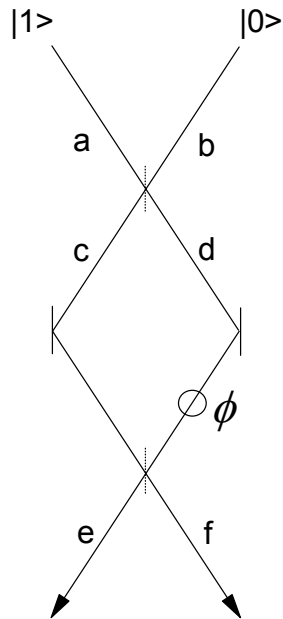

Figure 5. Interferometer arrangement used to implement the conditional photon transfer of Figure 4. A photon initially present in path a will be transferred to path f with probability amplitude $2iRT$ if $\phi = 0^0$, while it will exit with certainty in path e if $\phi = 180^0$.

The value of $\phi$ can be set to one of these two values using a controlled sign gate [1], where the control qubit is located in the y mode from the previous conditional transfer (or the x mode with a fixed $180^0$ phase shift). Alternatively, the controlled sign gate can be implemented using the controlled-NOT of Ref. 14 and two Hadamards, which gives a larger probability of success. The choice of the appropriate values of $R$ and $T$ will allow each of the successive transfers to occur with the desired probability amplitude, provided that the previous transfer occurred.

The correct value of $P_1$ corresponds to the sum over all of the states with one photon in the first mode of x, regardless of the occupation of the other modes:

$$P_1 = \sum_{j=1}^{n} f^2(j) \tag{8}$$

Here $f(j)$, a real number, is either a constant in the original KLM approach [1] or chosen to maximize the fidelity as described in Ref. [2].

The required value of $P_2$ corresponds to the total probability of all states having a photon in mode 2 of x, given that mode 1 is occupied. The fact that mode 1 is assumed to be occupied requires the introduction of a normalized probability amplitude $f'(j)$ given by

$$f'(j) = \frac{f(j)}{(\sum_{j'=1}^{n} f(j')^2)^{1/2}} \tag{9}$$

The necessary value of $P_2$ is then given by

$$P_2 = \sum_{j=2}^{n} f'(j)^2 \tag{10}$$

This process is then repeated for the remaining modes of x, which populates each term in the superposition with the correct probability amplitude and leaves the system in the entangled state of Eq. (1), as desired. An identical procedure is also applied to registers x' and y' to prepare them in the same entangled state.

The generation of two sets of entangled states of this kind requires a total of $2(n-1)$ controlled sign gates, since the first transfer in each register is not conditional.

## C. Entanglement of the two sets of registers

The final step in the creation of the entangled ancilla is the application of a factor of $(-1)^{jj'}$, which entangles the two sets of registers as required in Eq. (2). Two different procedures can be used, depending on the magnitude of $n$.

If $n$ is relatively small, the most efficient way to apply this phase shift is to perform $n^2$ controlled sign gates between each pair of qubits in x and x', with the x qubit the control and the x' qubit the target [1]. If we look at any given term in Eq. (2) with specific values of $j$ and $j'$, it can be seen that a minus sign will be applied a total of $jj'$ times, as required.

For large values of $n$, it is more efficient to avoid the quadratic dependence on $n$ by measuring the parities of registers x and x'. This is based on the observation that $jj'$ will be even if either $j$ or $j'$ is even, whereas it will be odd only if both $j$ and $j'$ are odd. The parity of register x can be obtained by introducing an additional ancilla qubit $q_a$ that is initially in the state 0, and then successively applying a CNOT gate between $q_a$ and every qubit in x, with the x qubits the control and $q_a$ the target. The ancilla will then have the value 0 if x has even parity and 1 if x has odd parity. The parity of x' can be determined in the same way using a second ancilla $q_b$.

Having determined the parities of x and x', we introduce a third ancilla $q_c$ initially in the state 0 and apply a controlled-controlled-NOT gate (Toffoli gate) to $q_c$ with $q_a$ and $q_b$ as the controls. (This operation can be performed using CNOTs and single-qubit operations.) Ancilla $q_c$ will now have the value 1 if and only if both $j$ and $j'$ are odd, so that the factor of $(-1)^{jj'}$ can now be applied by using a controlled sign gate applied to qubit 1 of registers x and y with $q_c$ as the control.

In order to avoid any entanglement between the terms in Eq. (2) and the states of the ancilla $q_a$, $q_b$, and $q_c$, it is also necessary to return all of these ancilla to the state $|0\rangle$. This can be done by repeating the controlled-controlled-NOT a second time, applying n CNOTs between $q_a$ and the qubits of x, and applying n CNOTs between $q_b$ and the qubits of x'. Including these operations, a total of $4n$ CNOT operations are required to apply the factor of $(-1)^{jj}$ in the limit of large n.

### D. Overall efficiency

Using $|\Phi^+\rangle$ states as a resource, controlled-NOT gates can be implemented using polarization techniques [14] with a probability of success of ¼. Controlled phase gates can be implemented directly using an interferometric approach, with a probability of success of 1/16 [1, 16]. It is more efficient, however, to implement the controlled phase gates using two Hadamards and a polarization-based CNOT [14], which allows the controlled sign gates to be implemented with a probability of success of ¼ as well.

For the case of $n = 3$, $2(n-1) = 4$ conditional transfers using a controlled sign gate are required to create the independent ancilla of Eq. (1). An additional $n^2 = 9$ controlled sign gates are required to apply the factor of $(-1)^{jj'}$. Thus the probability $P_S(3)$ of successfully generating an $n = 3$ set of entangled ancilla is given by

$$P_S(3) = \left(\frac{1}{4}\right)^{13}. \tag{11}$$

In the limit of large n, $2(n-1)$ controlled sign gates are again required to create the independent ancilla, followed by $4n$ controlled-NOTs to apply the factor of $(-1)^{jj}$. The corresponding probability of success is given by

$$P_S(n) = \left(\frac{1}{4}\right)^{6n-2} \qquad (n \gg 1) \qquad (12)$$

The fixed number of operations required for the controlled-controlled-NOT gates are insignificant in the limit of large n and have not been included in Eq. (12).

It can be seen that the number of attempts required to successfully generate the required ancilla state using post-selection increases exponentially with the value of $n$ and is relatively large even for the case of $n = 3$. We have not proven that this method is optimal, however, and the possibility remains that there may be more efficient methods, especially for small values of $n$.

The feasibility of post-selection approaches of this kind will depend on the availability of highly-efficient single-photon sources and detectors. Aside from the fact that the procedure must be repeated many times in order to successfully generate the desired entangled state, errors will be produced whenever a single-photon source fails to deliver one and only one photon, or when a detector fails to detect a photon. Since the total number of operations is on the order of $6n$, the probability of an error occurring in the preparation of the entangled state will be unacceptably high unless the error rates in the single-photon sources and detectors are correspondingly small.

## IV.    GENERATION OF ANCILLA USING QUANTUM WELLS

In this Section, we consider the possibility of generating the required ancilla states using a series of quantum wells [28, 29] and the Coulomb blockade effect [3, 4]. It will be found that the number of steps required to generate the ancilla scales as $n^2$ rather than exponentially.

The basic structure of interest is shown in Figure 6 for the case of $n = 3$, although the method is easily generalized to larger values of $n$. Each qubit in registers x and y are represented by a quantum dot labeled $D_1$ through $D_6$ in the figure. A thermal reservoir containing a large number of electrons is located on the right-hand-side of the array of quantum dots. The reservoir and each of the dots are separated by a tunnel barrier whose height can be controlled by applying an electrostatic potential to a series of gates labeled $B$ in the figure. The quantized energy levels in each of the dots can be controlled using another set of gates labeled U, which would be placed above or below the dots to produce a uniform electrostatic potential.

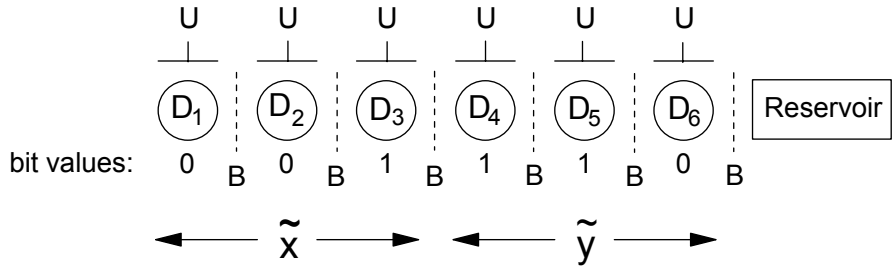

Figure 6. Generation of a photonic ancilla state by first generating the corresponding entangled state of $n$ electrons in a series of quantum dots labeled $D_1$ through $D_6$. The B gates are used to control the height of a tunneling barrier between the quantum dots, while the U gates are used to control the energies of the quantized states within each dot. A typical set of bit values is also shown below each quantum dot.

The goal is to first create an entangled state of $2n$ electrons as given in Eq. (2), where the presence of an excited electron will represent a logical value of 1 while the absence of an electron will represent the logical value 0. The electrons will be assumed to be in an excited state that allows the emission of a single photon into a single-mode optical fiber, so that the entangled state can be transferred to a photonic state if the coupling is sufficiently efficient. (The actual states of interest may be excitons, depending on the implementation.) In order to avoid the emission of photons during the time that the electronic state is being prepared, it may be necessary to first set up the electronic state in a set of quantum dots that are not coupled to the fibers and then simultaneously transfer all of the electrons to another set of quantum dots (not shown in the figure) that are coupled to the fiber modes.

The registers x and y have been rotated through an angle of $180^0$ compared to Figure 1, so that all of the 1s in the x register are now on its right-hand-side, while all of the 1s in the y register are now on its left-hand-side; the rotated registers have been denoted by $\tilde{x}$ and $\tilde{y}$. As a result of this rotation, all of the 1s are in adjacent locations and will form a contiguous block of $n$ excited electrons after the electronic state has been prepared, as illustrated by the bit values shown in Fig. 6.

The first step is to put all of the quantum dots into the logical state 0, which can be accomplished by raising their electrostatic potential and putting them into thermal equilibrium with the reservoir. A single excited electron can then be transferred from the reservoir to quantum dot $D_6$ using standard techniques in which the height of the tunnel barrier is reduced and the two regions are coupled for a time interval $\Delta t$ chosen to produce a complete Rabi oscillation. The transfer of two electrons in this process is prohibited by energy conservation, as has been demonstrated experimentally. Similar Rabi oscillations can then be used to transfer a single electron into registers $D_4$ and $D_5$. At this point, all of the qubits in register $\tilde{y}$ have the value 1, in analogy with Fig. 3.

Once the $\tilde{y}$ register has been filled with single electrons, the next step in the process is to adjust the height of the tunnel barrier between quantum dots $D_3$ and $D_4$ to perform a partial Rabi oscillation. This interaction is applied for a time interval $\Delta t'$ chosen in such a way that an electron will be transferred between the two quantum dots with a probability $P_1$ as specified in Eq. (8). If the transfer occurred, we need to perform a sequence of complete Rabi oscillations between successive quantum dots in register $\tilde{y}$ in order to move the missing electron (logical 0) all the way to the right, as illustrated in Figure 6. Since the first transfer may or may not have occurred, this process must be applied in any event and may therefore involve Rabi oscillations between two quantum dots that both contain an electron. In that case, the system is never left with two electrons in one of the dots and zero in the other because that is prohibited by the Coulomb blockade effect (energy conservation). The entire process is then repeated with probability $P_2$ and so forth until all of the terms in the superposition state of Eq. (1) have been generated with the appropriate probability amplitudes.

Similar techniques are also used to generate the same entangled state in a second set of rotated registers $\tilde{x}'$ and $\tilde{y}'$ (not shown). The factor of $(-1)^{jj'}$ can then be applied using the Coulomb interaction between the electrons in register $\tilde{x}$ and those in register $\tilde{x}'$. We assume that the electrons in registers $\tilde{y}$ and $\tilde{y}'$ are shielded with grounded conductors so that there is no interaction between their two sets of charges. We also note that the Coulomb interaction between the electrons within register $\tilde{x}$ itself will produce an undesired phase shift, but that depends only on the value of j and can therefore be cancelled out by applying an appropriate set of potentials to the U gates; the same can be done for register $\tilde{x}'$. With these precautions, the net electostatic interaction is proportional to the product of the sum of the charges on $\tilde{x}$ and $\tilde{x}'$, which is proportional to $jj'$. (This assumes that the distance between the two registers is much larger than the separation of quantum dots within a register.) Under these conditions, the factor of $(-1)^{jj'}$ will be automatically produced by the electrostatic interaction after an appropriate time interval.

The feasibility of an approach of this kind will depend on the rate of decoherence of the electrons as well as the efficiency with which the photons can be coupled into single-mode optical fibers. Although these are challenging problems, somewhat similar approaches are being considered for a variety of applications, including quantum computing [28-30]. Our results suggest the possibility of a hybrid approach that may be able to combine the potential advantages of both solid-state and linear-optics techniques. This may be especially useful for intermediate-range applications in quantum communications [31-33].

## V.     SUMMARY

We have discussed two different methods for the generation of the entangled ancilla photon states required for linear optics quantum computing. Ancilla states with $n$ photons can be generated in a straightforward way using elementary linear optics gates and post-selection. Given elementary CNOT gates that succeed with a probability of ¼ [14, 18, 19], this approach generates the correct ancilla state with a probability of $(1/4)^{6n-2}$ in the limit of large $n$. We have not shown that this method is optimal, however, and more efficient post-selection methods may eventually be found, especially for small values of $n$. This method can be used to generate the ancilla required for either the original KLM approach or the high-fidelity approach, and the same resources are required in either case.

We also suggested a possible solid-state approach for generating the entangled ancilla photon states. The corresponding entangled state of $n$ electrons would first be generated using a series of quantum dots and tunnel barriers. The nonlinearity required for this process comes from the Coulomb blockade effect, which prevents more than one electron from being transferred to a single quantum dot. After the entangled state of the electrons has been generated, it would be transferred to $n$ single-mode optical fibers using a set of efficient optical couplers [5].

The feasibility of post-selection techniques will depend on the availability of high-efficiency single-photon sources and detectors, while the feasibility of the solid-state approach may be limited by decoherence of the electrons or inefficient coupling into single-mode optical fibers. Advances continue to be made in all of these areas and the generation of the required entangled ancilla states may become feasible at some time in the future. Nevertheless, the difficulties involved in either of these techniques suggest that it may be desirable to consider the possibility of alternative approaches that minimize or eliminate the need for large numbers of ancilla photons and high-efficiency detectors [6].

We would like to acknowledge valuable discussions with Jonathan Dowling, Michael Fitch and Todd Pittman. This work was supported in part by funding from ARO, ARDA, ONR, NSA, and IR&D.